\documentclass[a4paper,twocolumn,11pt,accepted=2022-01-10]{quantumarticle}
\pdfoutput=1
\usepackage[utf8]{inputenc}
\usepackage{amsmath}
\usepackage{amsfonts}
\usepackage{amssymb}
\usepackage{bm}
\usepackage{graphicx}
\usepackage{xcolor}
\usepackage{microtype}
\usepackage{ dsfont }

\usepackage[numbers,sort&compress]{natbib}

\begin{document}
\title{The quantum annealing gap and quench dynamics in the exact cover problem
}
\author{Bernhard Irsigler}
\author{Tobias Grass}
\affiliation{ICFO-Institut de Ciencies Fotoniques, The Barcelona Institute of Science and Technology, 08860 Castelldefels (Barcelona), Spain}

\begin{abstract}
Quenching and annealing are extreme opposites in the time evolution of a quantum system: Annealing explores equilibrium phases of a Hamiltonian with slowly changing parameters and can be exploited as a tool for solving complex optimization problems.  In contrast, quenches are sudden changes of the Hamiltonian, producing a non-equilibrium situation. Here, we investigate the relation between the two cases. Specifically, we show that the minimum of the annealing gap, which is an important bottleneck of quantum annealing algorithms, can be revealed from a dynamical quench parameter which describes the dynamical quantum state after the quench. Combined with statistical tools including the training of a neural network, the relation between quench and annealing dynamics can be exploited to reproduce the full functional behavior of the annealing gap from the quench data. We show that the partial or full knowledge about the annealing gap which can be gained in this way can be used to design optimized quantum annealing protocols with a practical time-to-solution benefit. Our results are obtained from simulating random Ising Hamiltonians, representing hard-to-solve instances of the exact cover problem.
\end{abstract}
\let\clearpage\relax\maketitle

\section{Introduction}
Quantum annealers \cite{amin09,Albash2018,Hauke2020} are an emerging technology which combine features from analog quantum simulators \cite{lewenstein2012ultracold,Blatt2012,Aspuru-Guzik2012} and digital, gate-based quantum computers \cite{Preskill2018}. Like in quantum simulators, the engineering of a Hamiltonian is the working scheme in quantum annealers, but these devices feature also microscopic design opportunities which allows one to choose a Hamiltonian such that its ground state solves a specific instance of a general computational problem. For example, any NP-hard  optimization problem can be mapped onto an Ising spin model with tunable couplings \cite{barahona}. Together with a transverse field term producing quantum fluctuations, this Ising Hamiltonian is the central element of existing quantum annealers, specifically the commercial D-wave device \cite{dwave11}, as well as in atomic annealing architectures \cite{hauke15,grass16,glaetzle17,torggler17,xingze20} which have been proposed.
In the absence of decoherence, the quantum annealing algorithm tries to find the ground state of the model through an adiabatic dynamical process, i.e.~adiabatic state preparation or adiabatic quantum computing \cite{Farhi2001}. In the presence of decoherence, the dynamical process to find the ground state resembles more a conventional thermal cooling process, yet with a quantum feature which provides control over quantum fluctuations \cite{Kadowaki1998,brooke99}.
In both cases, the dynamical evolution encounters configurations in which the energy gap above the instantaneous ground state becomes exponentially small  \cite{joerg08,young10,joerg10,altshuler10,knysh16},  and this becomes the bottleneck of the quantum annealing method. Therefore, different strategies to avoid small energy gaps have been reported, including inhomogeneous transverse fields \cite{dickson11,farhi11,dickson12,lanting17}, non-stoquastic driver Hamiltonians \cite{seki12,crosson14,hormozi17,albash19,oxfidan19}, reverse or biased annealing \cite{Perdomo-Ortiz2011,ohkuwa18,baldwin18,grass19}, as well as combinations thereof \cite{tang21}. There are also different hybrid algorithms which may avoid the annealing bottleneck \cite{chancellor17,grass17,morley19,callison19}.

A simple strategy which can improve the fidelity of the quantum annealing method is the deceleration of the annealing process around the minimal gap. In this way, the efficiency of the adiabatic Grover algorithm could be improved from $T\sim \mathcal{N}$ to $T\sim \sqrt{\mathcal{N}}$ \cite{Roland2002}, where $T$ is the annealing time and $\mathcal{N}$ is the Hilbert space dimension of the system. This remarkable result, however, relies on the analytical accessibility of the problem. Strikingly, also in the case of quantum annealing far away from the adiabatic regime, experiments on the D-wave device have shown that pausing the evolution near the minimal gap is beneficial for the annealing fidelity \cite{Marshall2019}. This experimental observation has theoretically been explained in Ref.~\cite{Chen2020}, attributing the improvement of the quantum annealing to thermalization. No clear time-to-solution improvement from pausing, however, has been found yet \cite{Chen2020}. 
Nevertheless, the pausing feature suggests that, in both adiabatic and non-adiabatic scenarios, knowledge of the instantaneous energy gap above the ground state could help to speed-up the annealing process. In the present work, we investigate how such knowledge can be obtained through simple quench experiments, and be refined through machine-learning techniques.

Although adiabatic evolution and quenching are two opposite extremes with a clear discrepancy in the time-dependence of the field, relations between the two approaches have been studied \cite{callison21,brady21}. A physical phenomenon which can occur after a quantum quench is the so-called dynamical quantum phase transition, which may be manifested from the behavior of an order parameter after the quench \cite{yuzbashyan06,sciolla10}. Specifically, it has been shown for the long-times dynamics in long-range transverse-field Ising models that the non-zero magnetization of a system which is initially in the ferromagnetic phase averages to zero during the evolution after quenching with a sufficiently strong transverse field \cite{Halimeh2017a,Halimeh2017b,Homrighausen2017,Zunkovic2018,Jafari2019,Mishra2020}. This phenomenon has also been experimentally observed in a trapped-ion experiment \cite{zhang17}. 

The present manuscript considers a somewhat similar scenario, but with some ``randomness'' in the Ising couplings. In the absence of a transverse field, this ``randomness'' makes the system behave rather like a spin glass than a ferromagnet. In fact, we will define the ``random'' couplings through randomly generated (but well characterized) instances of an NP-hard optimization problem, specifically the exact cover problem which is frequently studied  in the context quantum annealing, see Refs.~\cite{Farhi2001,altshuler10,grass19}. Analyzing the dynamical behavior of different order parameters, related to either the local magnetization of the system in the glassy ground state
or to the paramagnetic order in a strong transverse field, we first determine a ``critical'' field strength in the quench dynamics. Assuming that the dynamical behavior in the quench dynamics is related to the equilibrium paramagnetic and glassy states, we establish through numerical simulations that the annealing bottleneck coincides with the critical field value of the quench. We then show that this connection between quench and annealing dynamics can be used to design quantum annealing protocols with a significant fidelity enhancement, as compared to the standard annealing protocol with a constant ramp speed. Importantly, the proposed method to improve the annealing result does not rely on any previous knowledge of the problem instance, but only on information which can be collected through a few quench experiments.

In a second step, we explore how the connection between quench and annealing dynamics can be extended beyond the matching at only one critical field strength. To this end, we have trained a neural network which maps the order parameter as a function of the quench field to the energy gap during the annealing process. We note that, in comparison to reinforcement learning strategies which in the past have been applied to problems from optimal control \cite{bukov18} and quantum annealing \cite{lin20}, our approach has only a small amount of training requirements, as our method is based on supervised training of  only a single-layer network. Nevertheless, after this training process, it becomes possible to fully and almost exactly reproduce from the quench data the annealing gap as a function of the transverse field strength. With this information, it then becomes possible to adjust the instantaneous ramp speed of the annealing protocol. Not surprisingly, this yields a further fidelity improvement, as compared to our first method which had not required any network training.

Our paper is organized in three Sections: In Sec. II, we describe the studied computational problem, and define the annealing gap, as well as the order parameters in the quench dynamics. We distinguish two types of quenches: switching the transverse field on or off. In Sec. III, we present the results which in subsection (A) consists of a relation between quenched order parameters and annealing gap, and in subsection (B) of an application of this connection to the quantum annealing protocol. In both subsections, we further distinguish between (1.) the connection which can immediately be made at the critical field strength,
and (2.)  the connection at arbitrary field strengths, obtained after training a neural network. Finally, in Sec. IV, we summarize our results, and briefly discuss the technological advances associated with our results.

\section{Definitions}
\subsection{Model}
We define the annealing Hamiltonian as
\begin{equation}
\hat{H}[h(t)]= (1-h(t))\hat{H}_p + h(t)\hat{H}_q,
\label{annealHam}
\end{equation}
herein, we introduced the problem Hamiltonian $\hat{H}_p$, the fluctuation Hamiltonian $\hat{H}_q$, and the time-dependent field  $h(t)$ {which varies monotonously increasing from 0 to 1}.  We focus on the exact cover (EC) problem to be embedded in the problem Hamiltonian. In the language of spins, the EC problem is described by a set of clauses $C$ containing $M$ tuples of randomly selected spins, each point either up or down. Here, we restrict ourselves to the EC3 problem, i.e., those tuples will be triples.  In EC3, a clause is fulfilled if exactly two of the three spins point up while one of them points down.  Since, different clauses can share the same spins, the problem can become hard to solve. It is assumed that for $M\approx N$ the problem is hard, where $N$ is the number of spins in the system. In this regime, many instances have exactly one solution, dubbed as unique satisfying assignment. We focus solely on those instances. The classical EC3 problem can be expressed in a Hamiltonian in the following way:
\begin{equation}
\hat{H}_p=J\sum_{(i,j,k)\in C} (\hat{\sigma}^z_i+\hat{\sigma}^z_j+\hat{\sigma}^z_k-1)^2,
\label{ec3Ham}
\end{equation}
where $J$ is the energy scale. The set $C$ contains $M$ random triples, dubbed clauses, such that
\begin{equation}
C=\{(i,j,k):i\neq j\neq k \in [1,2,..N]\},
\label{clauses}
\end{equation}
where $N$ is the number of spins in the system. If a spin configuration satisfies all clauses in $C$, its energy associated with Eq.~\eqref{ec3Ham} is zero. On the other hand, every clause which is not fulfilled by a spin configuration leads to an energy cost of at least $4J$. Therefore, for the instances of $C$ which have unique satisfying assignment, there is one zero-energy state, being the non-degenerate ground state of $\hat{H}_p$.

For the  fluctuation Hamiltonian we employ a transverse spin term as 
\begin{equation}
\hat{H}_q=J\sum_{i=1}^N\hat{\sigma}^x_i.
\label{flucHam}
\end{equation}
Here, we define the fluctuation Hamiltonian using the same energy scale $J$ as in the problem Hamiltonian, but in the annealing Hamiltonian, as described by Eq.~(\ref{annealHam}), the dimensionless function $h(t)$ controls the energetic weight of $\hat{H}_q$ and $\hat{H}_p$.

Throughout this manuscript, we make use of the quspin package for the exact diagonalization, quenches and time evolutions \cite{Weinberg2017,Weinberg2019}. All codes are available at Ref.~\cite{Irsigler2021}.

\subsection{Annealing gap}

We define the annealing gap $\Delta(h)$ as the energy gap between the two lowest energy levels $E_n(h)$ of the Hamiltonian in Eq.~\eqref{annealHam} rescaled by the bandwidth $W(h)$ for time-independent $h(t)=h$:
\begin{equation}
\begin{split}
\Delta(h)&=\frac{E_1(h)-E_0(h)}{W(h)},\\
W(h)&=E_{2^N}(h)-E_0(h).
\end{split}
\label{gap}
\end{equation}
Here, $E_0$ and $E_1$ denote the energy of ground state and the first exited state, respectively.

\begin{figure}[ht]
\centering
\includegraphics[width=.82\columnwidth]{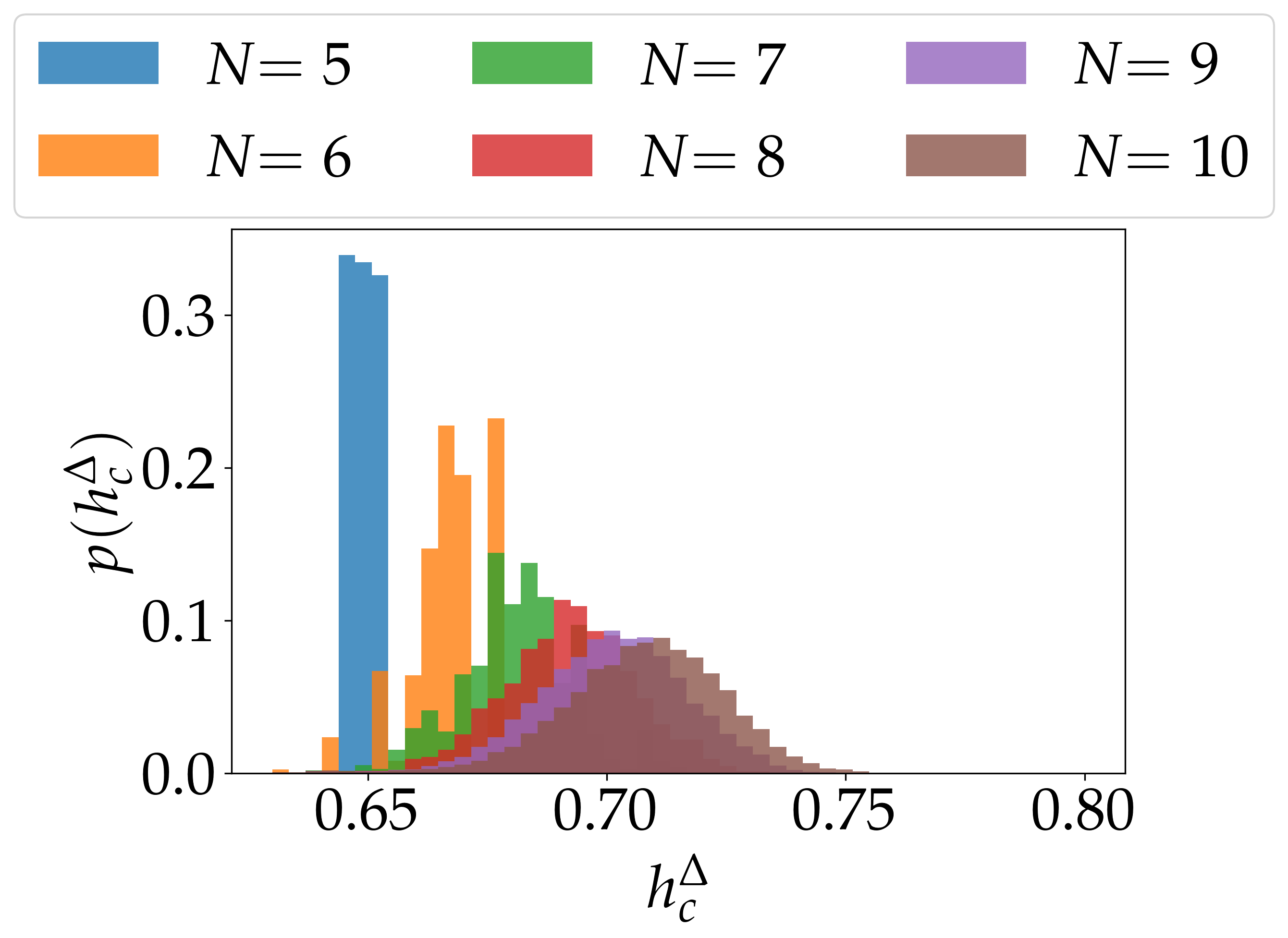}
\caption{Probability distribution of the {critical field $h_c^\Delta$ for which the annealing gap is minimal} for different system sizes $N$. }
\label{minGap}
\end{figure}

We define the critical field $h_c^\Delta = \mathrm{argmin}(\Delta(h))$ at which the annealing gap is minimal. We compute $h_c^\Delta$ by determining numerically the global minimum of the gap of the annealing Hamiltonian in Eq.~\eqref{annealHam} for a time-independent field $h$ with $0<h<1$. The result is shown in Fig.~\ref{minGap} as a probability distribution. We observe that the probability distributions become broader with increasing $N$ as more diverse configurations of clauses, see Eq.~\eqref{clauses}, can emerge.

\subsection{Quench dynamics}
{In the thermodynamic limit, \textit{the dynamical quantum phase transition} is achieved through a sudden quench of the field $h(t)$ in the Hamiltonian in Eq.~\eqref{annealHam}, and the dynamical phase transition is characterized through appropriately chosen order parameters \cite{yuzbashyan06,sciolla10}. We borrow these concepts for our study of finite systems to define \textit{dynamical quench parameters} (DQPs), in full analogy to the  order parameters in the thermodynamic limit. Assuming values between 0 and 1, these DQPs characterize the dynamical properties of different quantum states within the quench dynamics.}  

{In the following, we differentiate between the instantaneous ground states $|\psi_{h}\rangle$, which are eigenstates of the time-independent Hamiltonian in Eq.~\eqref{annealHam} with $h(t)=h$, and the time-evolved states $|\psi_{h(t)}(t)\rangle=\hat{U}_{h(t)}(t,0)|\psi_{h_i}\rangle$ where $h_i$ is the initial field.  Here, 
\begin{equation}
\hat{U}_{h(t)}(t,0)=\mathcal{T}\exp\left[-i\int_0^t\hat{H}[h(t')]dt'\right]
\label{timeEvolution}
\end{equation}
}is the time-evolution operator. Herein, $\mathcal{T}$ is the time-ordering operator and $\hbar=1$.
In the following, we treat two different kinds of quenches associated with different initial states and different DQPs. 
\subsubsection{``switch on''} 
\label{switch_on}
This procedure is similar to the one used in Ref.~\cite{Zunkovic2018} for treating the homogeneous Ising model, {i.e., spin-spin interactions between two spins only depends on their spatial distance.} 
The initial state is  the ground state of $\hat{H}[h=0]$ which is then quenched to the final field $h_f$ with finite value $0<h_f\leq1$. The associated observable is 
\begin{equation}
G(h_f,t)= \frac{1}{N}\sum_{i=1}^N\langle\psi_0|\sigma^z_i|\psi_0\rangle\langle\psi_{h_f}(t)|\hat{\sigma}^z_i|\psi_{h_f}(t)\rangle.
\end{equation}

We note that for $t=0$,  $|\psi_{h_f}(t)\rangle=|\psi_0\rangle$, and therefore $G(h_f,0)=1$.
The corresponding DQP is defined as
\begin{equation}
G(h_f)=\lim_{\tau\rightarrow\infty}\frac{1}{\tau}\int_0^\tau G(h_f,t) dt.
\label{sigmaG}
\end{equation}
In practice, of course, the limit of $\tau \rightarrow \infty$ is replaced by a finite time $\tau$ for which $G(h)$ has converged sufficiently well to its actual long-time limit.  To this end, we have determined $G(h)$ for different values of $\tau J =10, 20, 50, 100$. As we found the deviations to be much smaller than the actual value of $G(h)$, we have chosen the value $\tau J=20$ for our numerical investigations.

\subsubsection{``switch off''} 
\label{switch_off}
This procedure is more relevant for quantum annealing since here the fluctuation field is initially maximal, $h=1$, and is then quenched to {$h_f$ with $0\leq h_f<1$}. Also, it does not require any prior {knowledge about the initial ground state $|\psi_0\rangle$ of the problem Hamiltonian.} For the ``switch off'' procedure the associated observable is defined as
\begin{equation}
X(h_f,t) = \frac{1}{N}\sum_{i=1}^N\langle\psi_{h_f}(t)|\hat{\sigma}^x_i|\psi_{h_f}(t)\rangle
\end{equation}
As in in the example above,  for $t=0$,  $|\psi_{h_f}(t)\rangle=|\psi_1\rangle$ and therefore $X(h_f,0)=1$.
The corresponding DQP is defined as
\begin{equation}
X(h_f)=\lim_{\tau\rightarrow\infty}\frac{1}{\tau}\int_0^\tau X(h_f,t) dt.
\label{sigmaX}
\end{equation}
{As for $G(h_f)$, also here, in practice, the limit $\tau\rightarrow \infty$ has to be replaced by a sufficiently large time scale.}

\begin{figure}[ht]
\centering
\includegraphics[width=.52\columnwidth]{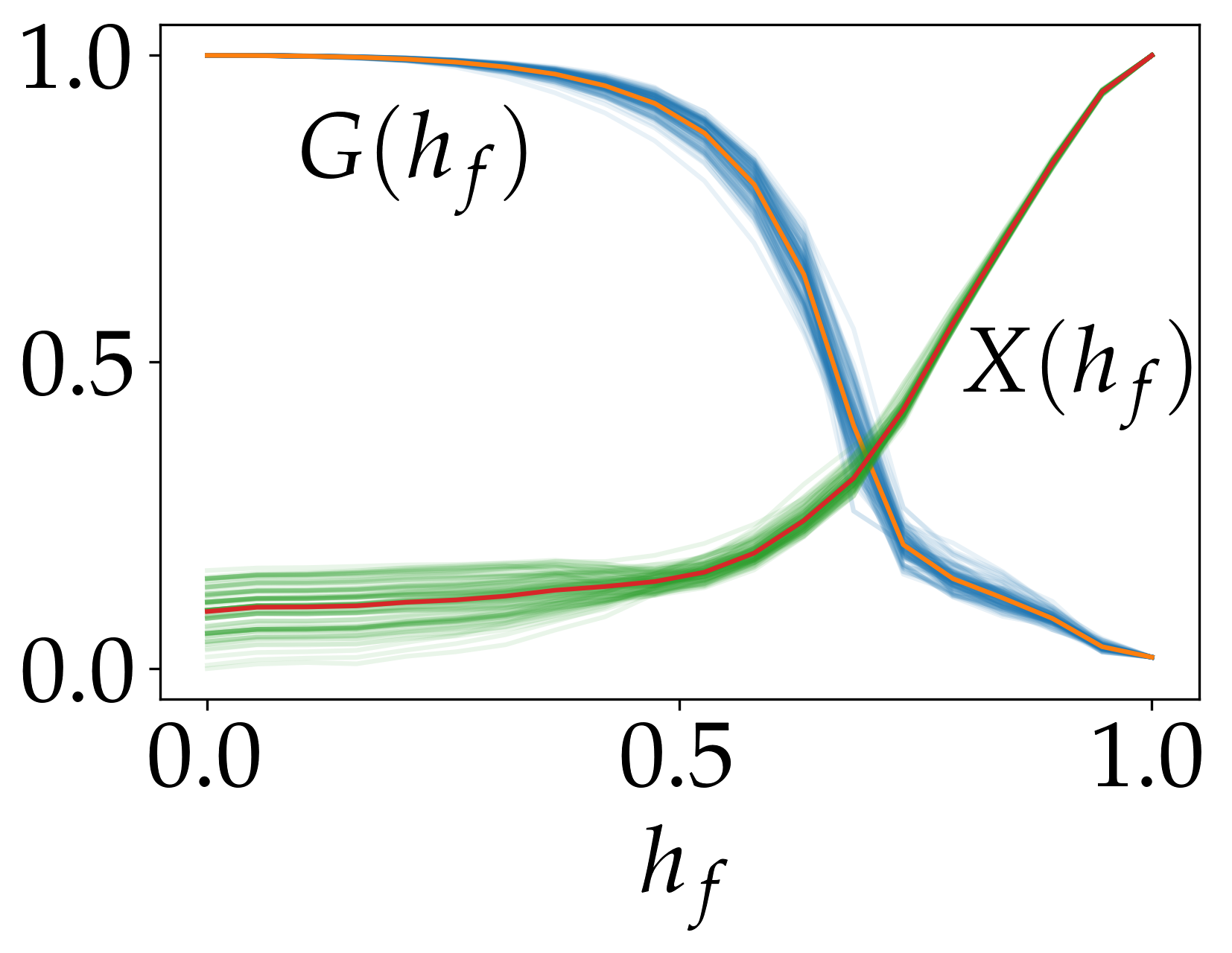}
\caption{DQPs as a function of the quench field $h_f$ for 100 different instances of the problem Hamiltonian in Eq.~\eqref{ec3Ham} with $N=10$. Thick lines correspond to the respective averages.}
\label{DQP}
\end{figure}

In Fig.~\ref{DQP}, we present the DQPs as they appear from the numerical quenches. Therein, $G(h_f)$ and $X(h_f)$, defined in Eqs.~\eqref{sigmaG} and \eqref{sigmaX}, respectively, are plotted for 100 different instances of the problem Hamiltonian with $N=10$.

\section{Results}
{We will now report our results of numerical experiments in which, as described in the previous Section, the EC3 problem Hamiltonian and a transverse magnetic field term are quenched with respect to each other. Our goal is to use the data obtained in this way in order to optimize quantum annealing protocols. Therefore, we proceed in the following two-fold way: First, in Sec. 3.1, we show that there exists a sizable correlation between the minimal annealing gap and the DQP. This result establishes that the two very different, dynamical approaches are indeed connected. It provides the possibility to predict the annealing gap from the result of quench experiments, either via linear regression (Sec. 3.1.1) or using a properly trained neural network (Sec. 3.1.2). While the linear regression approach is sufficient to obtain an estimate of the position $h_c^\Delta$ of the \textit{minimal} gap, the neural network approach is able to reproduce well the \textit{full} annealing gap $\Delta(h)$ as a function of the field $h$. Second, in Sec. 3.2, we exploit this predictability of the gap in order to gain a time-to-solution improvement over the standard annealing procedure. We distinguish between protocols which are restricted to knowledge about the position of the minimum gap (Sec. 3.2.1), and protcols which make use of the full gap function (Sec. 3.2.2).}

\subsection{Relation between annealing gap and quench dynamics}
Annealing and quenching are at the opposite sides in terms of time dependence, but they may actually reveal the same information about a quantum state. In the following, we investigate how the annealing gap can be estimated from the information drawn from the DQP. {To this end, we have simulated the quench dynamics using exact diagonalization and calculated the DPQs, as described in the previous Section, and we have calculated the low part of the energy spectrum of the system as a function of $h$}.
 
\subsubsection{Critical field from {multivariate} linear regression}
{In this Section, we employ a multivariate linear regression in order to determine the critical field $h^\Delta_c$ at which the annealing gap $\Delta(h)$ is minimal from the DQPs. 
To this end, we have discretized the quench field $h_f$ into 20 points $h_f^n$, for $n=1,2,...20$, with $h_f^0=0$ and $h_f^{20}=1$. The values which the DQP assumes at these points, e.g., $G(h_f^n)$, are then used in the multivariate regression. This means that we assume the following functional relation:
\begin{equation}
    \mathbf{h}_c^\Delta = \beta_0\mathbf{1}
    + \sum_{n=0}^{20} \beta_n \mathbf{G}(h_f^n).
\end{equation}
Here, the bold symbols denote vectors containing the 100 instances of Fig.~\ref{DQP} as elements. The 21 parameters $\beta_n$ determine the linear contribution of the different $\mathbf{G}(h_f^n)$ to the critical field $\mathbf{h}_c^\Delta$. The $100\times21$ matrix
\begin{equation}
    A = [\mathbf{1},\mathbf{G}(h_f^1),\mathbf{G}(h_f^2),\dots,\mathbf{G}(h_f^{20})]
\end{equation}
is called design matrix. Together with the parameter vector $\vec{\beta}=(\beta_0,\beta_1,\dots,\beta_{20})^T$ a least-squares estimator $||A\vec{\beta}-\mathbf{h}_c^\Delta||^2$ can be employed which can be minimized with respect to $\vec{\beta}$ by standard routines.
}

\begin{figure}[ht]
\centering
\includegraphics[width=.85\columnwidth]{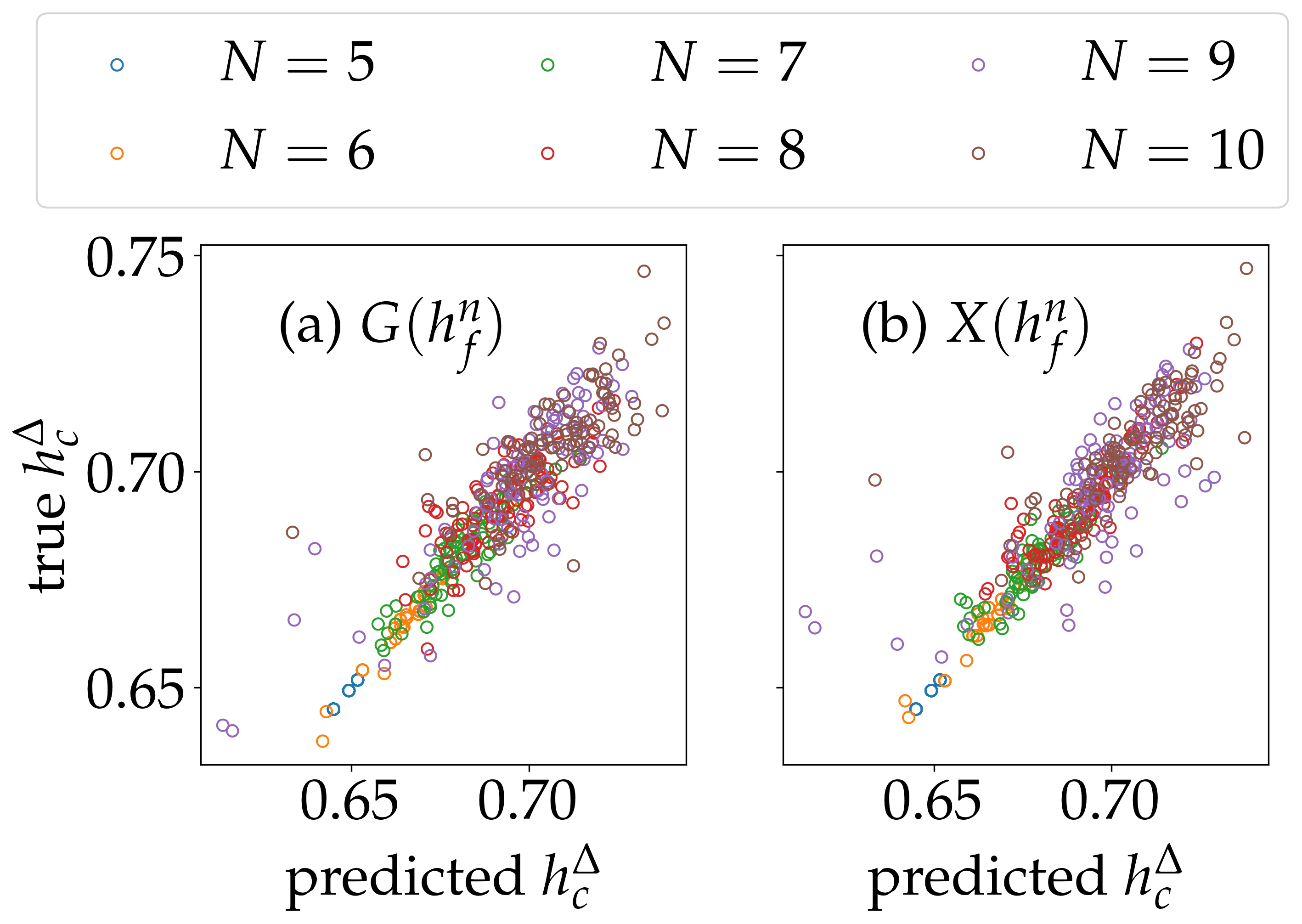}
\caption{Predicted values for the field of the minimal gap $h_c^\Delta$ through multivariate regression of the (a) $G(h_f^n)$ and (b) $X(h_f^n)$ estimators, respectively, versus true values of $h_c^\Delta$.}
\label{predMinGap}
\end{figure}

\begin{figure}[ht]
\centering
\includegraphics[width=.52\columnwidth]{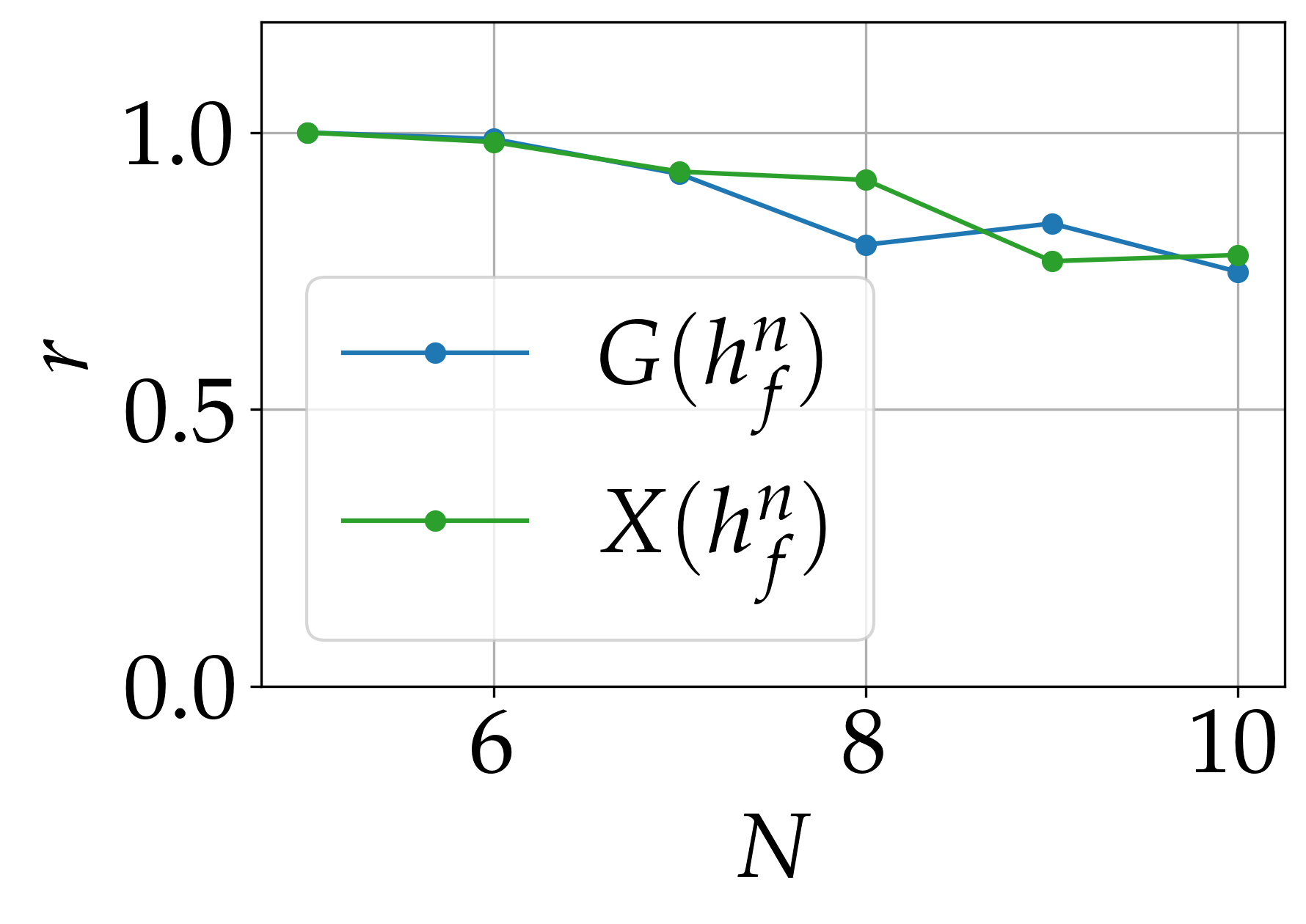}
\caption{The Pearson's $r$, {computed from the data presented in Fig.~\ref{predMinGap} using Eq.~\eqref{pearson_r}, as a function of the system size $N$ for both quench approaches introduced in Sec.~\ref{switch_on} and \ref{switch_off}, respectively.}}
\label{pearson}
\end{figure}

The predictions from the $G(h_f^n)$ are presented in Fig.~\ref{predMinGap}(a) and those of the  $X(h_f^n)$ are presented in Fig.~\ref{predMinGap}(b). The predictions are both plotted versus the true values of the minimal gap.
In order to estimate the quality of the prediction, we introduce the Pearson's $r$ as a measure for the linear correlation between to random variables $x$ and $y$:
\begin{equation}
r=\frac{\left\langle (x - \langle x\rangle)(y - \langle y\rangle)\right\rangle}{\sqrt{\langle(x - \langle x\rangle)^2\rangle}\sqrt{\langle(y - \langle y\rangle)^2\rangle}}
\label{pearson_r}
\end{equation}
In particular, we want to quantify how well our predictive methods work by interpreting $x$ as the true values and $y$ as the predicted values {of $h_c^\Delta$ in Fig.~\ref{predMinGap}.  In Fig.~\ref{pearson}, we show the Pearson's $r$ quantifying the predictive power of the $G(h_f^n)$ and $X(h_f^n)$ as a function of the number of spins in the system.  The fact that the system size dependence is found to be non-monotonic makes it difficult to estimate the scaling with $N$. Although a tendency of decreasing with the system size can be seen, the correlation between the predicted and the true minimal gap remain strong ($r\gtrsim 0.75$) thoughout the considered range of $N$. We conclude that the DQP contains information about the quantum state change during the annealing process.}

\subsubsection{Full gap function from neural network \label{IIA2}}
In the previous section, we have shown that the value for the critical field at the minimal gap $h_c^\Delta$ can be obtained from the DQP as a function of the quench field $h_f$. In the present section, we employ modern machine-learning techniques \cite{Mehta2019} to acquire not only the value of $h_c^\Delta$ but the full function $\Delta(h)$, see Eq.~\eqref{gap}, in the entire regime $0<h<1$.

Compared to conventional algorithms, neural networks work best for a large amount of data. Therefore, we restrict ourselves to the system size of $N=7$ as well as the data from $X(h_f)$ only and create 5000 random instances of the Hamiltonian in  Eq.~\eqref{ec3Ham}. We then compute $\Delta(h)$ as well as $X(h_f)$, see Eqs.~\eqref{gap} and \eqref{sigmaX}, respectively. Both functions are evaluated on an equidistant grid of 128 points. Note that it is not required by the algorithm that both functions are discretized by the same number of data points.

We make use of the established high-level package Keras \cite{keras} in order to construct a simple neural network with one single dense layer exhibiting a linear activation function.  We use the mean square error as the loss function, a batch size of 64 and the ADAM optimizer \cite{Mehta2019}. Out of the full data set of 5000 instances, we use 4000 as the train/validation data set and 1000 as the test data set. The validation data set is another split of 20\% of the 4000 instances.

\begin{figure}[ht]
\centering
\includegraphics[width=.61\columnwidth]{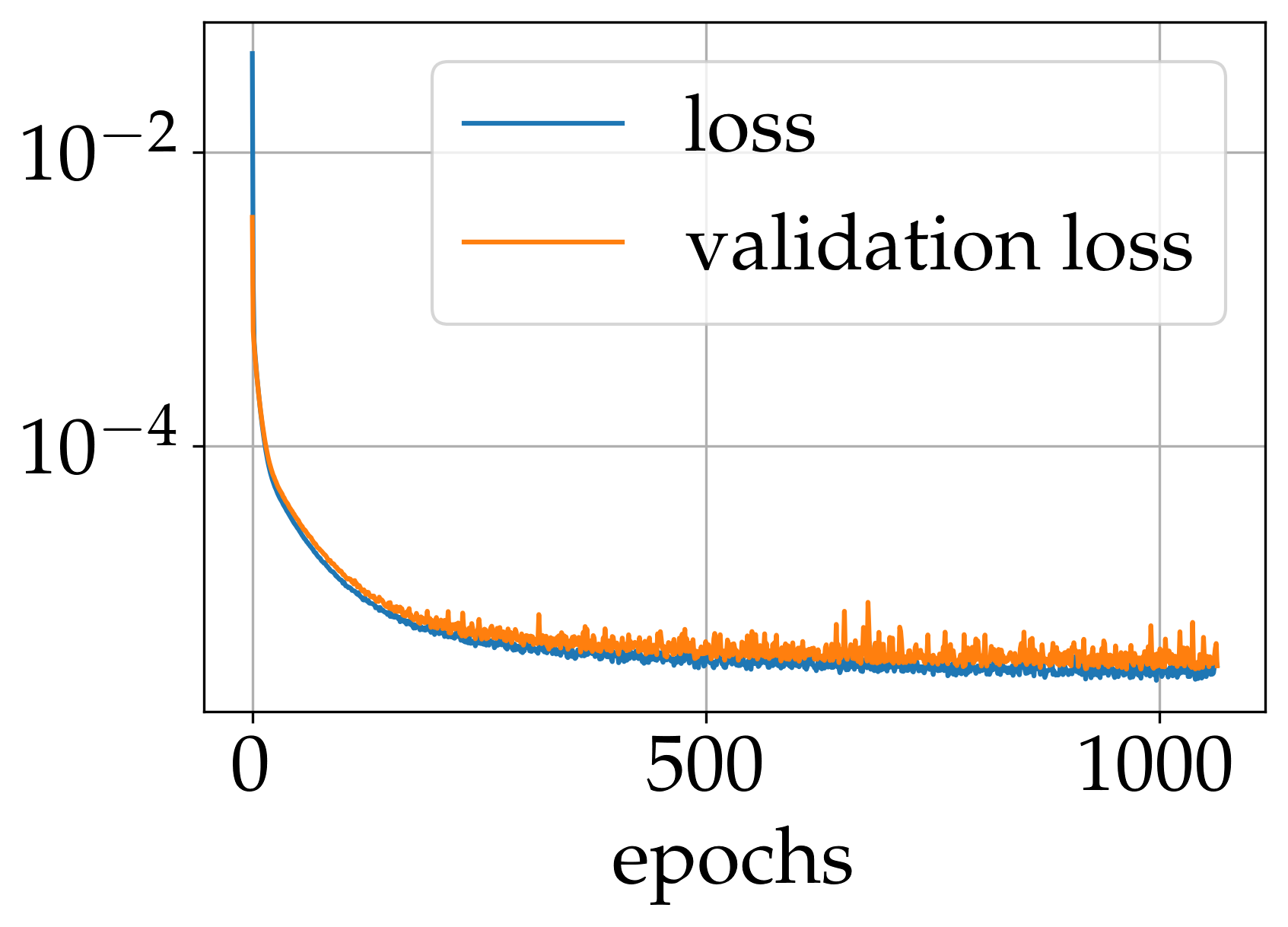}
\caption{Training of the neural network: evolution of the loss and the validation loss over the episodes. We plot the validation loss in order to rule out possible overfitting.}
\label{train}
\end{figure}

We show the training of the neural network in Fig.~\ref{train}.  After 600 epochs, we achieve a loss of  $\sim10^{-6}$. In order to make sure that there is no overfitting occurring during the training, we plot the loss for the validation set, i.e.,  the loss on the data the neural network has not seen before.  Since validation loss behaves exactly like the loss function itself,  we can be certain that there is no overfitting happening.  The only free parameter in our neural network is the learning rate which enters in the optimizer. We have found through a grid search that $5.5\cdot 10^{-4}$ leads to the lowest loss.

\begin{figure}[ht]
\centering
\includegraphics[width=.94\columnwidth]{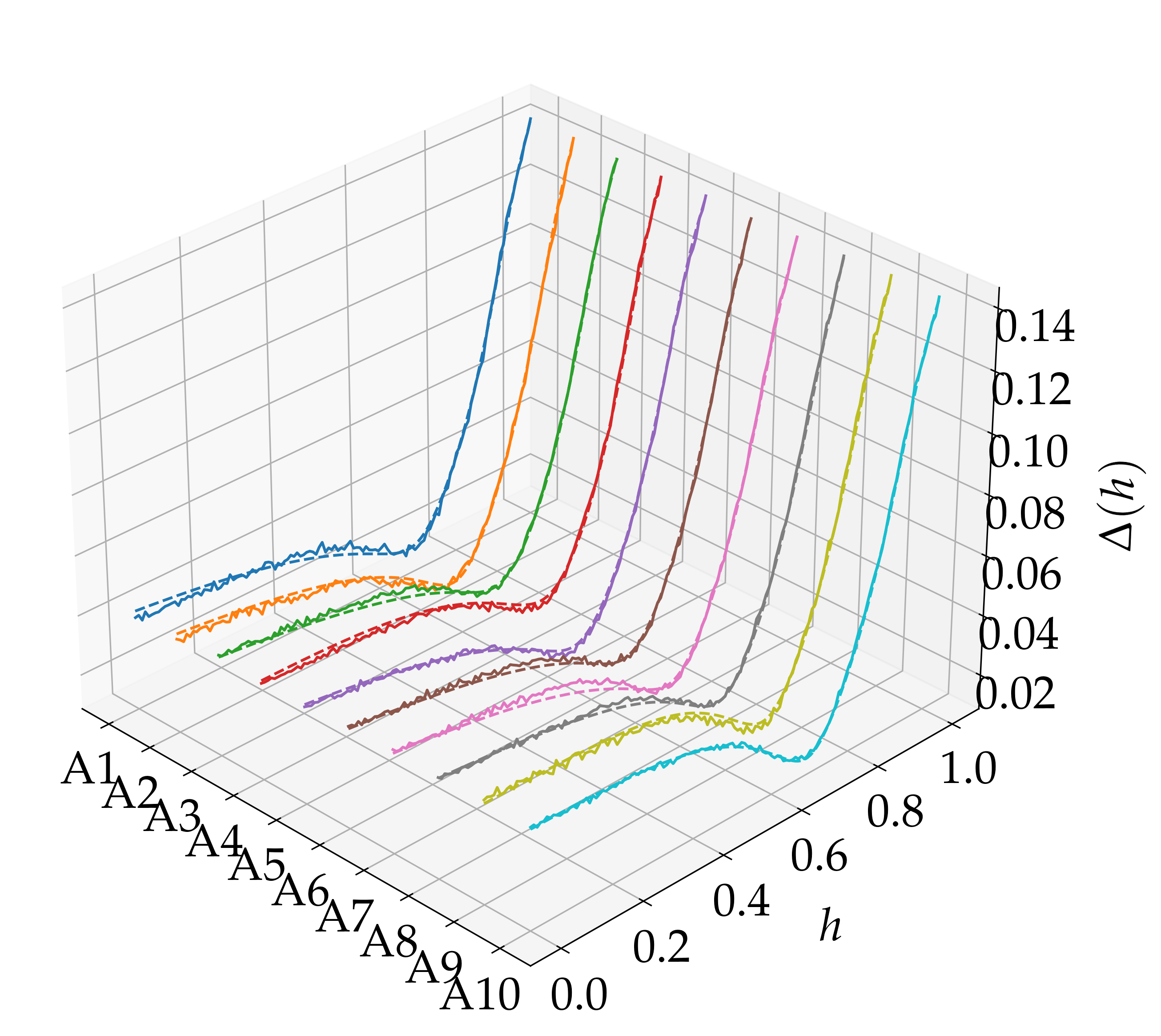}
\caption{Predictions for the gap of the test data as solid lines and the respective true gap functions as dashed lines. 
The different instances are labeled by A1 to A10. The diagonal of the respective Hamiltonian $\hat{H}_p$, see Eq.~\eqref{ec3Ham}, is provided in the supplemental material \cite{Irsigler2021}.}
\label{prediction}
\end{figure}

Finally, we plot some instances for the  gap function as predicted by the trained neural network from the test data set in Fig.~\ref{prediction}. We compare these predictions with the true gap functions out of the test set shown as dashed curves in Fig.~\ref{prediction}.  

\subsection{Time-to-solution improvement for annealing}
In this section, we want to draw a technical benefit out of the knowledge of the annealing gap. Specifically, we design annealing protocols with non-linear time dependencies which, at a fixed annealing time, lead to a better final fidelity than the protocol with homogeneous ramp speed.
For the construction of these inhomogeneous protocols, we first restrict ourselves to exploiting the correlation between minimal gap and the DQP. As shown in the previous Section, this information is readily obtained from quench experiments. In a second step, we will further refine the annealing strategy by exploiting the full knowledge of the annealing gap, which we have shown to be accessible through a combination of quench data and machine learning.

\subsubsection{Protocols with slow-down at critical field strength}
We perform calculations for the full time evolution in order to simulate the annealing procedure at finite times. In this way, we {compare linear annealing protocols, i.e., with a constant ramp speed}, and annealing protocols {with a varying ramp speed.}
The linear protocol has the following form
\begin{equation}
h_{\text{lin}}(t) = 1-\frac{t}{T},
\label{linProt}
\end{equation}
where $T$ is the duration of the annealing procedure.  {For the protocol with variable ramp speed, we choose} a combination of two quadratic functions
\begin{equation}
h_{\text{quad}}(t,t_c)=
\begin{cases}
1+b_1t+c_1t^2, \text{ if }t<t_c\\
a_2+b_2t+c_2t^2, \text{ if }t\geq t_c.
\end{cases}
\label{quadProt}
\end{equation}
Here, we have introduced a time $t_c$ at which the annealing protocol should slow down. Therefore, we impose the condition of a vanishing velocity at $t_c$, i.e., $\partial_t h_{\text{quad}}(t,t_c)|_{t=t_c}=0$. Moreover, we demand that $h_{\text{quad}}(t,t_c)$ is a smooth function, and that it assumes the values $h_{\text{quad}}(t,t_c)=1$ for $t=0$ and $h_{\text{quad}}(t,t_c)=0$ at $t=T$. These conditions define the coefficients $b_1,c_1,a_2,b_2,$ and $c_2$ in Eq.~\eqref{quadProt}. Finally, with the knowledge of a field strength $h_c$ at which the protocol is supposed to slow down (ideally the field strength at which the gap becomes smallest), we are able to associate the time $t_c$ with $h_c$ via $t_c=(1-h_c)T$. In this way, the construction of the quadratic protocol incorporates a slow-down of the annealing speed in the vicinity of $h_c$, at fixed duration $T$ of the annealing procedure. With $t_c$ being determined by $h_c$, we denote the quadratic protocol by $h_{\text{quad}}(t,h_c)$.

\begin{figure}[ht]
\centering
\includegraphics[width=\columnwidth]{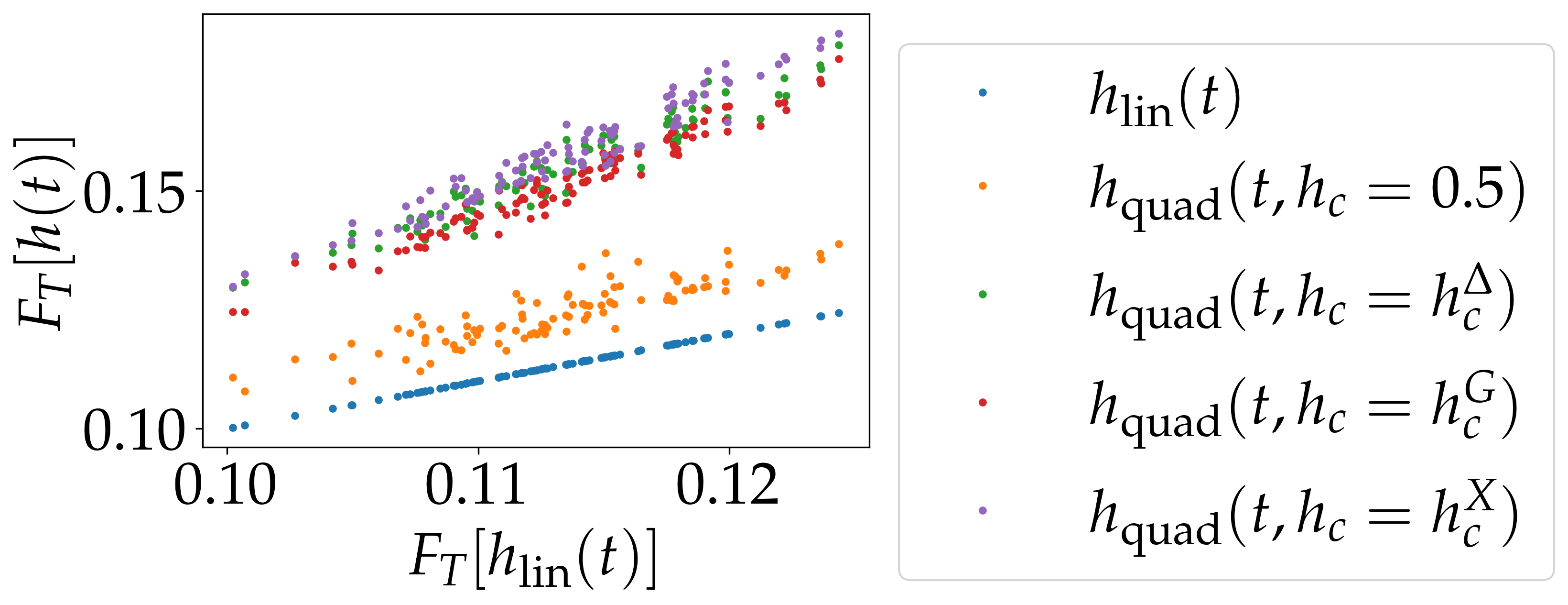}
\caption{Fidelity for five different annealing protocols for $N=10$ and $T=1$, see Eqs.~\eqref{linProt} and \eqref{quadProt}.}
\label{fidelity_N10}
\end{figure}

\begin{figure}[ht]
\centering
\includegraphics[width=.97\columnwidth]{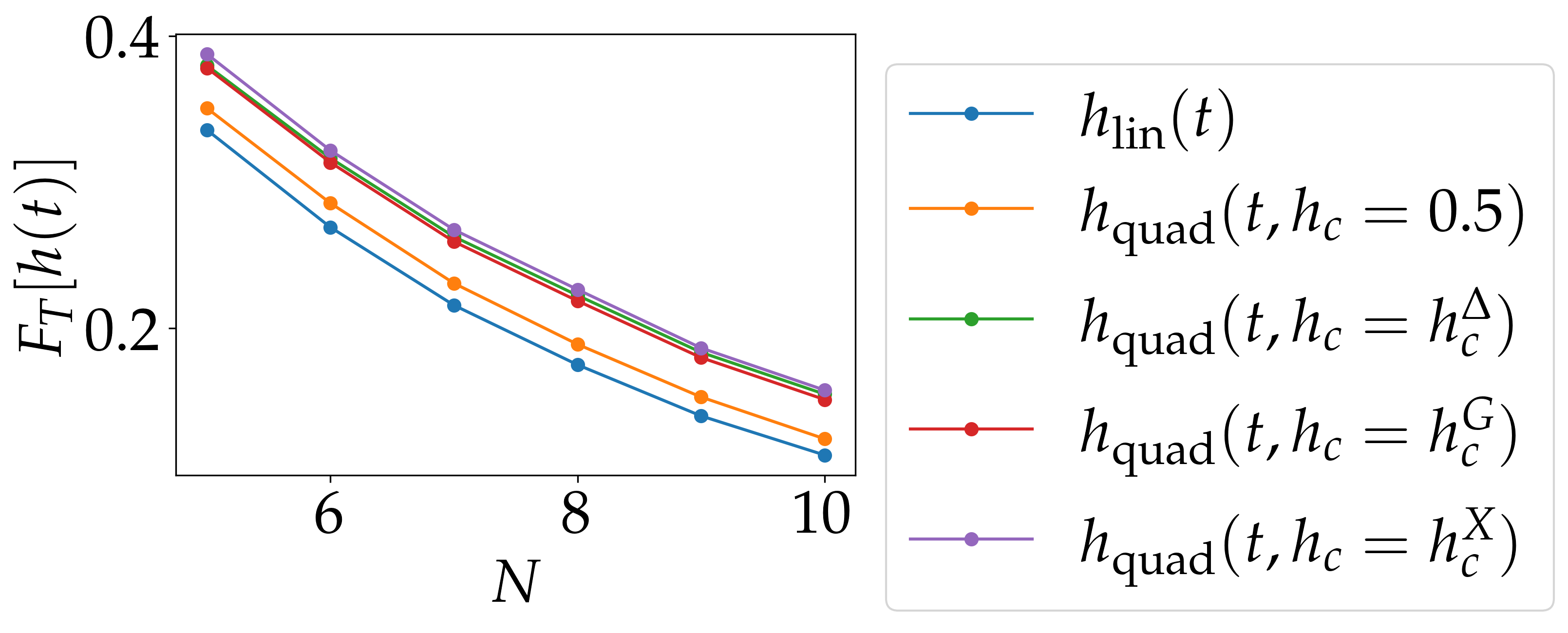}
\caption{Scaling of the average fidelity as a function of the number os spins in the system for $T=1$.}
\label{fidelityScaling}
\end{figure}

\begin{figure}[ht]
\centering
\includegraphics[width=.81\columnwidth]{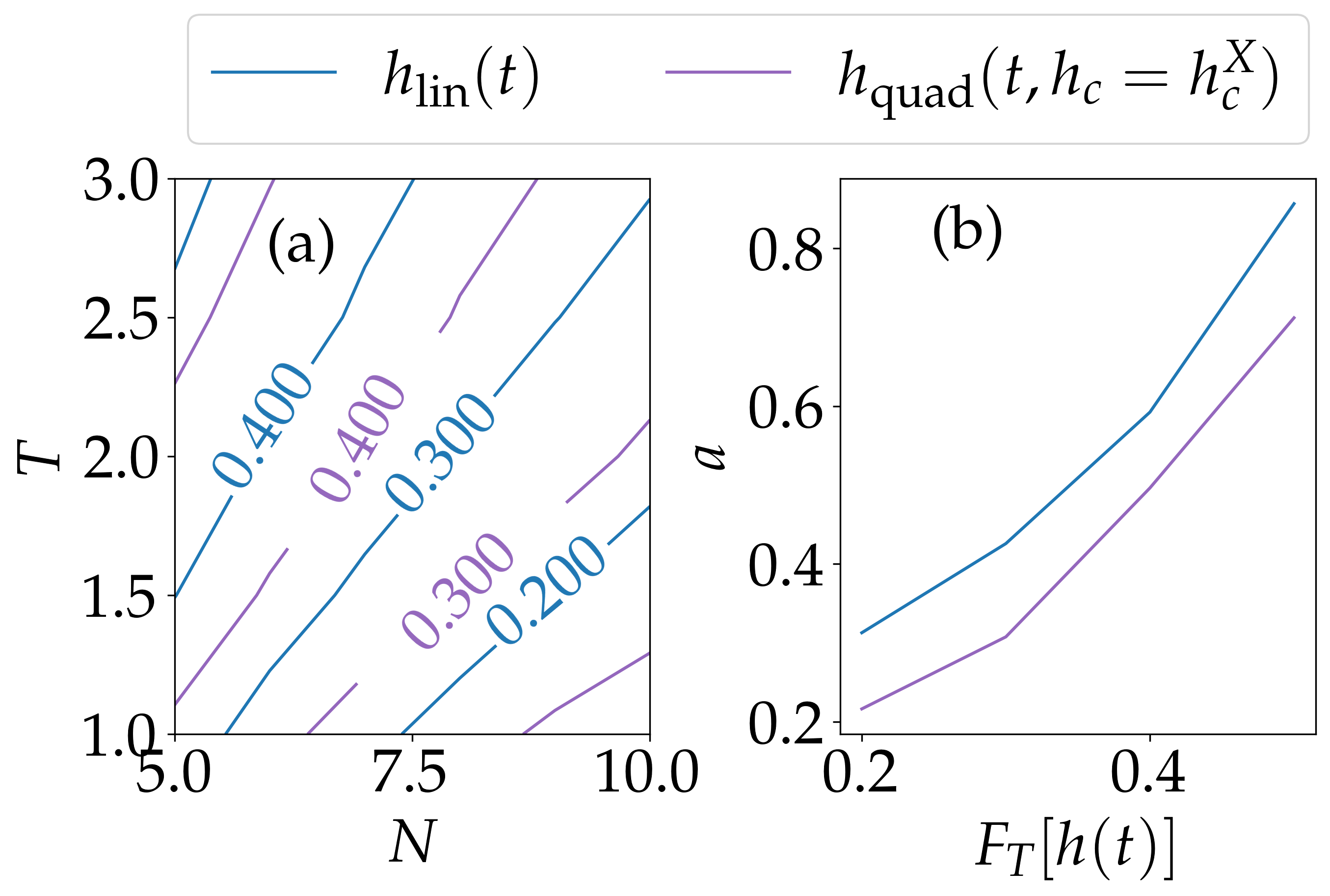}
\caption{{Scaling of the worst and the best annealing protocol shown through the isobaric lines of constant average fidelity in (a). Linear fitting coefficients of the scaling in (b).}}
\label{scaling}
\end{figure}

To evaluate the quality of the different protocols, we define the fidelity as the overlap
\begin{equation}
    F(t)=|\langle\psi_{h}|\psi_{h(t)}(t)\rangle|.
\end{equation} 
Herein, $|\psi_{h}\rangle$ is the instantaneous ground state of the Hamiltonian $H[h(t)=h]$ defined in Eq.~\eqref{annealHam}. The time-evolved state $|\psi_{h(t)}(t)\rangle$ is computed through the time evolution in Eq.~\eqref{timeEvolution} including the time variation of the protocol $h(t)$. Of course, the interest in this quantity is largest at the end of the annealing procedure, $t=T$, where the squared fidelity corresponds to the success rate of the annealer. As a functional of the protocol $h(t)$, we define the final fidelity $F_T[h(t)]=F(T)$, and analyze it for different protocols in Figs.~\ref{fidelity_N10}-\ref{alpha_stat}. 

In particular, we are interested how the choice of the slow-down field value $h_c$ affects the fidelity. The simplest choice of $h_c$ is to set it to a constant, e.g., $h_c=0.5$. This choice will serve as a reference. Of course, a better choice would be to use the instance-specific field strength corresponding to the minimal annealing gap as the critical field. We denote this choice by $h_c=h_c^\Delta$. However, this information is typically unknown from an experimental point of view. As established in the previous Section, data from quench experiments allows to make instance-specific guesses of the region with small gap, and therefore $h_c$ can be coupled to the previously defined order parameters $G$ and $X$ (see Eqs.~(\ref{sigmaG}) and \ref{sigmaX})). We define
\begin{align}
h_c^G, \text{ where }G(h_f=h_c^G)=0.5\\
h_c^X, \text{ where }X(h_f=h_c^X)=0.5.
\end{align}

In Fig.~\ref{fidelity_N10}, we show exemplarily the results for the fidelity of four different quadratic annealing protocols compared to the fidelity of the linear protocol $h_\text{lin}(t)$ defined in Eq.~\eqref{linProt}. While one quadratic protocol is chosen independent of the respective instance of the problem Hamiltonian, the other quadratic protocols depend on an instance-specific field value $h_c$ as an input parameter. We observe that these last three protocols yield a significant improvement with respect to the instance-independent protocol. Even though all these three protocols perform similarly well, it is curious to note that choosing the  parameter $h_c$ to be exactly at the minimal gap $h_c^\Delta$ is not the best choice. Instead, the choice of $h_c$ such that $X(h_c)=0.5$ is seen to outperform all other protocols for almost all instances. We note that with this choice, $h_c$ tends to be shifted towards slightly smaller values, i.e. the slowest annealing occurs slightly after the minimal gap has been passed.

We are now interested in the scaling of these protocols with respect to the number of spins in the system. The average fidelity as a function of the number of spins is shown in Fig.~\ref{fidelityScaling} for the five different protocols.
In order to quantify the time-to-solution improvement we compute the annealing time evolution for different annealing times $T$. We then determine the isobaric lines of constant average fidelity in order to find the scaling of $T$ as a function of $N$. {Those isobaric lines are shown in Fig.~\ref{scaling}(a) for the best and the worst protocol. We observe that both scale approximately linearly, $T\approx a N$, at least in the regime $5\leq N\leq 10$, with the linear coefficient $a$. By linearly fitting the isobaric lines of constant average fidelity, we determine the respective linear coefficient $a$ shown in Fig.~\ref{scaling}(b). We observe that the linear scaling coefficient $a$ of the linear annealing protocol and the quadratic protocol exhibit a similar functional behavior. However, the parameter $a$ of the linear protocol seems to be offset  by a constant, i.e., the linear protocol is always less efficient than the quadratic protocol.}

\subsubsection{Protocols utilizing the full gap function}

In the previous section, we have observed that the knowledge of a critical field obtained from, e.g., the DQP, can improve the time-to-solution of the annealing process. In this section, we will investigate whether we can improve this even further, assuming we have knowledge about the full gap function. As shown in Sec.~\ref{IIA2}, the combination of quench data and neural network training might be an accurate source of this information.

Inspired by the adiabatic theorem, we assume that the optimal protocol should scale with the size of the gap
\begin{equation}
\frac{dh_\mathrm{opt}(t,\alpha)}{dt}\propto|E_1(t)-E_0(t)|^\alpha.
\label{adiabatic}
\end{equation}
Here, we have introduced a free parameter $\alpha$ which in the most standard version of the adiabatic theorem \cite{Albash2018} would be $\alpha=2$. We note that the adiabatic theorem contains a matrix element of the Hamiltonian which couples the two states of interest. As it has been done in Ref.~\cite{Roland2002}, we approximate this matrix element by the time derivative of the protocol $h_\mathrm{opt}(t,\alpha)$. Furthermore, we note that Eq.~\eqref{adiabatic} considers only transitions between the energetically lowest two states. In fact, our systems of interest have $2^N$ states, and generally all transitions depleting the ground state contribute to the loss of annealing fidelity. Thus, Eq.~\eqref{adiabatic} is certainly to be understood as an approximation. This also means that there might not be a universal $\alpha$ which is optimal for all the spin glass instances we investigate.

We now use Eq.~\eqref{adiabatic} in  order to find an optimized annealing protocol. In a discretized form, this means
\begin{equation}
\Delta t_i = \frac{|E_1(t_i)-E_0(t_i)|^{-\alpha}}{\sum_j|E_1(t_j)-E_0(t_j)|^{-\alpha}}T,
\label{fullGap}
\end{equation}
where we have introduced the discretized time $t_{i+1}=t_i+\Delta t_i$. Also note that this protocol is normalized to the duration of the annealing process $\sum_i\Delta t_i =T$. For $\alpha=0$, the protocol becomes the standard linear protocol as Eq.~\eqref{linProt}.

\begin{figure}[ht]
\centering
\includegraphics[width=.64\columnwidth]{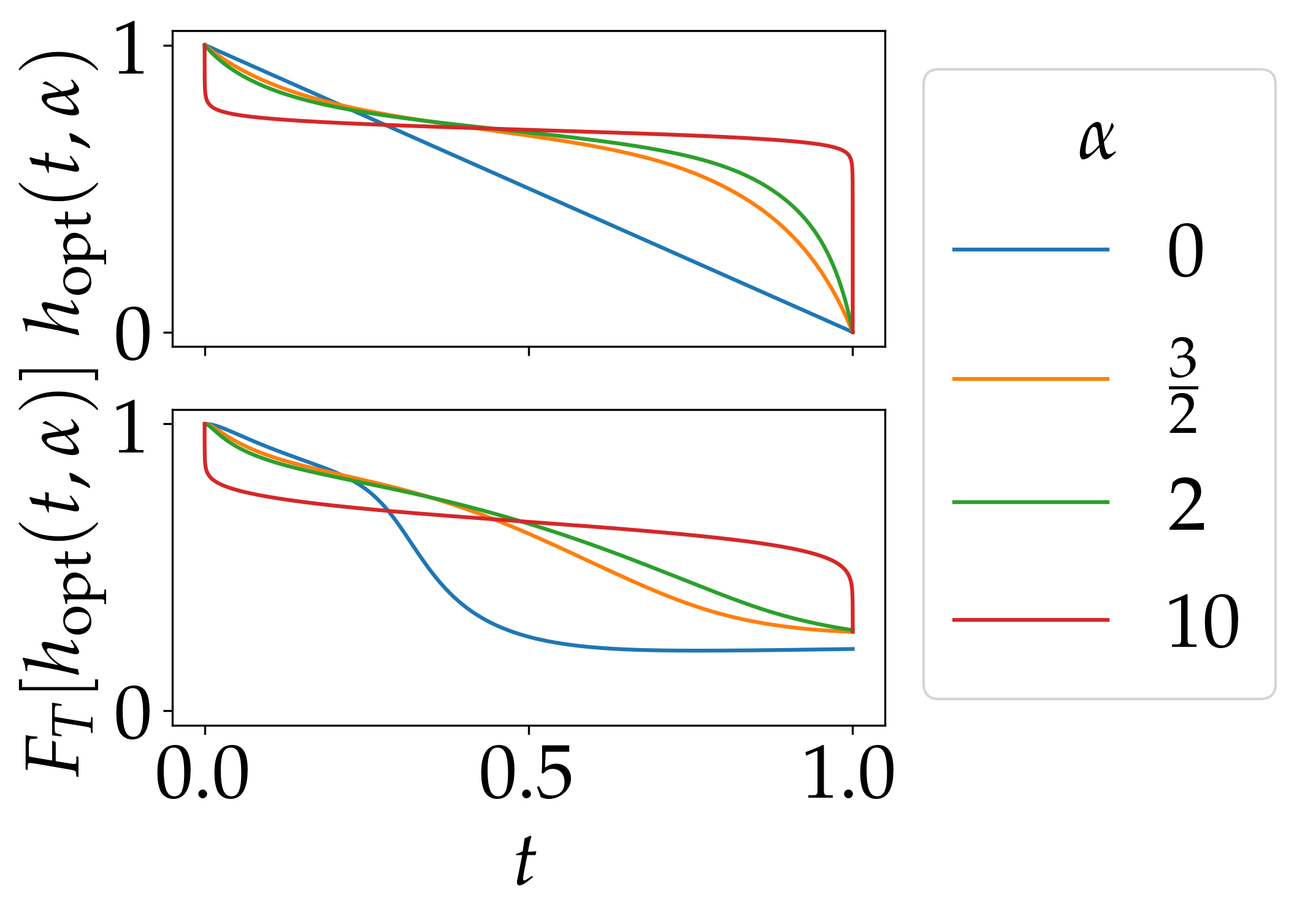}
\caption{Full gap annealing protocols according to Eq.~\eqref{fullGap}  for different values of $\alpha$ and a single Hamiltonian instance for $N=7$ and $T=1$. The diagonal of the respective Hamiltonian $\hat{H}_p$, see Eq.~\eqref{ec3Ham}, is provided in the supplemental material with the label B1 \cite{Irsigler2021}.}
\label{fidVst}
\end{figure}

\begin{figure}[bt]
\centering
\includegraphics[width=1\columnwidth]{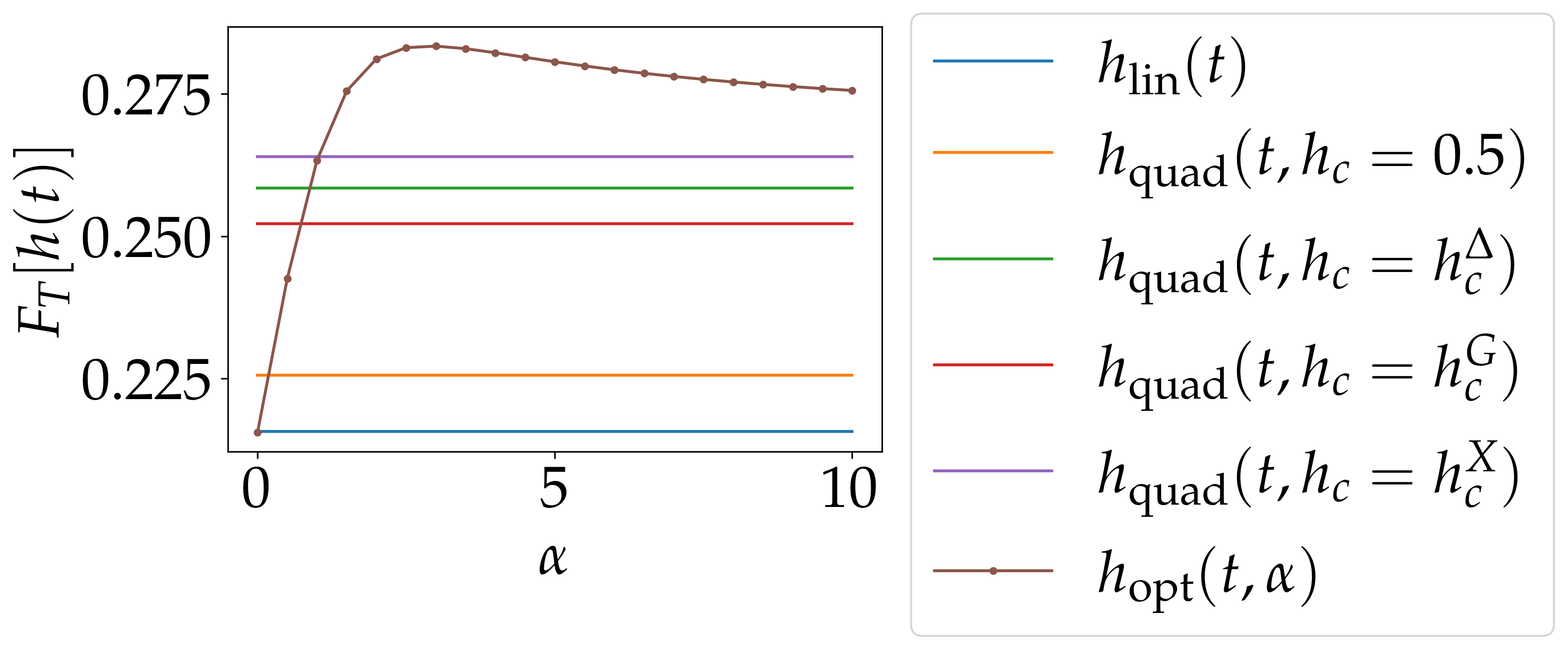}
\caption{Final fidelity as a function of $\alpha$ for a single instance. The diagonal of the respective Hamiltonian $\hat{H}_p$, see Eq.~\eqref{ec3Ham}, is provided in the supplemental material with the label B1 \cite{Irsigler2021}.}
\label{fidVsAlpha}
\end{figure}

In Fig.~\ref{fidVst}, we show annealing protocols $h_\mathrm{opt}(t,\alpha)$ constructed according to Eq.~\eqref{fullGap} for different values of $\alpha$ as a function of time. We also show the evolution of the fidelity with respect to time. First, the protocols $h_\mathrm{opt}(t,\alpha)$ with finite $\alpha$ improve the fidelity at the end of the annealing, as compared to the linear protocol $\alpha=0$. Secondly, we observe that the fidelity for larger values of $\alpha$ is similar at $t=T$. In Fig.~\ref{fidVsAlpha}, we show the fidelity at the end of the annealing process as a function of $\alpha$ for a single instance. We observe that the maximum is achieved for $\alpha=3$. {As we have mentioned before, and as our discussion in the next paragraph will make more explicit,} this value is not universal and differs from instance to instance. We also include in Fig.~\ref{fidVsAlpha} the values for the fidelity of the quadratic protocols investigated in the section before.  We conclude that knowledge of the full gap function can indeed lead to a further improvement of the annealing process.

\begin{figure}[ht]
\centering
\includegraphics[width=.69\columnwidth]{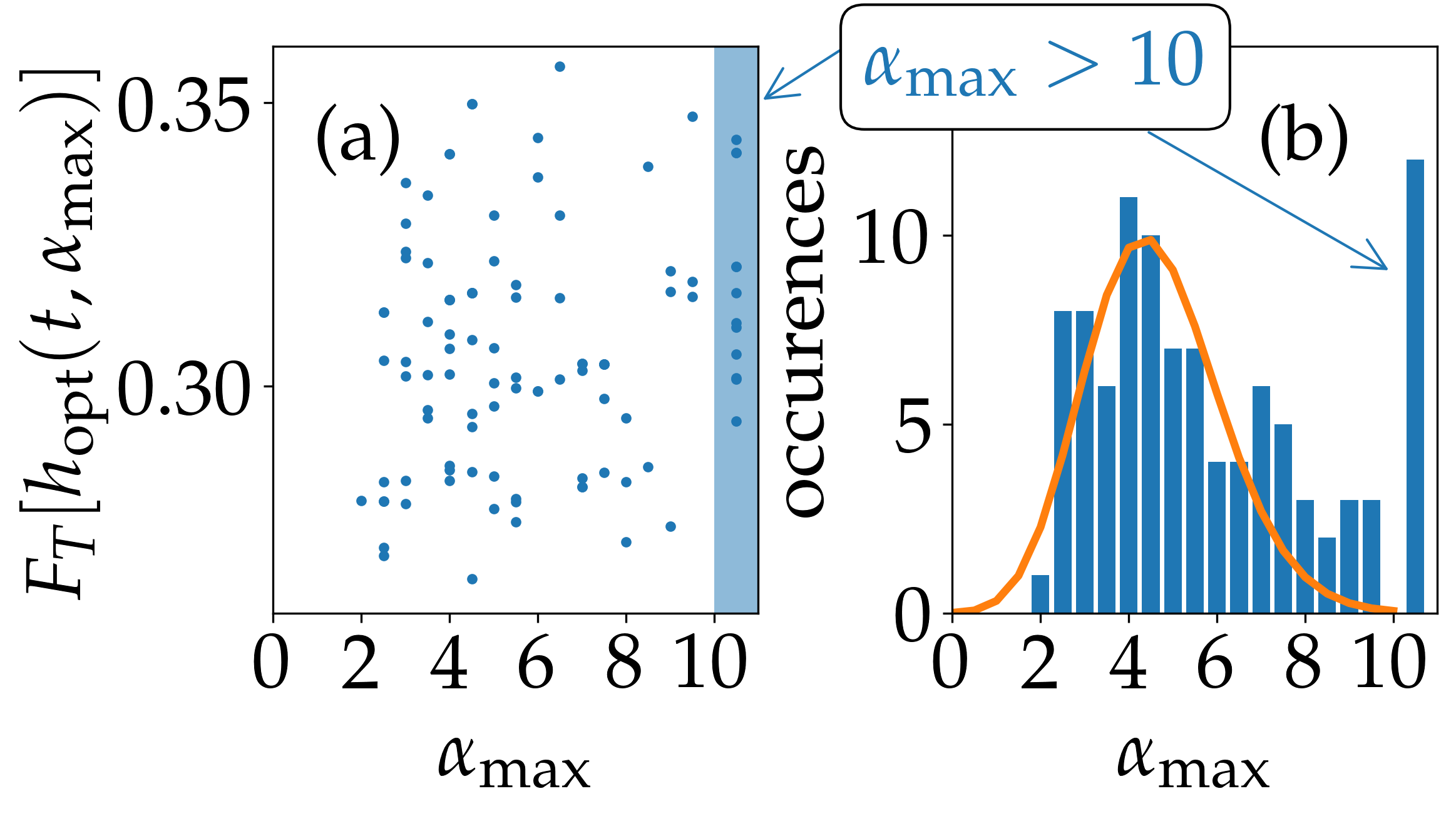}
\caption{{Statistics of $\alpha_\text{max}$ for which the fidelity is maximal: (a) fidelity at $\alpha_\text{max}$ versus $\alpha_\text{max}$, (b) histogram of $\alpha_\text{max}$. Values larger than 10 are grouped and highlighted. The orange curve is a Poisson distribution  fitted to the values $\alpha_\text{max}\leq10$.}}
\label{alpha_stat}
\end{figure}

The non-monotonic behavior of the annealing protocol $h_\mathrm{opt}(t,\alpha)$ in Fig.~\ref{fidVsAlpha} leads to the focus on the maximal reachable fidelity. We define the value $\alpha_\text{max}$ at which the fidelity for the protocol $h_\mathrm{opt}(t,\alpha)$ is maximal. To answer the question whether the maximum fidelity, i.e. the complexity of a certain instance, is correlated to $\alpha_\text{max}$, we consult Fig.~\ref{alpha_stat}(a). By plotting the data of 100 instances, we see no tendency to a correlation between $\alpha_\text{max}$ and the respective maximum fidelity $F_T[h_\mathrm{opt}(t,\alpha)]$. Moreover, we deploy a histogram for the values of $\alpha_\text{max}$ shown in Fig.~\ref{alpha_stat}(b). We fit a Poisson distribution $P_\lambda(\alpha_\text{max})$
to the data in the range  $0\leq\alpha_\text{max}\leq10$ leading to a parameter $\lambda\approx4.6$. In the Poisson distribution, this parameter at the same time represents the mean value as well as the variance. Remarkably, a mean value of 4.6 is much larger than the commonly encountered value of 2 in adiabatic theorems \cite{Albash2018}.

\section{Summary and Outlook}

We have investigated the annealing and quench dynamics of a spin model incorporating the exact cover problem. While the quench dynamics exhibits a jump of the  dynamical quench parameters (DQP), the annealing dynamics undergoes an equilibrium transition between a paramagnetic and a glassy state. We have shown that the  DQP can be used to determine the minimal annealing gap of the system through multivariate regression, see Figs.~\ref{predMinGap} and \ref{pearson}. Moreover, by means of a one-layer neural network, it even becomes possible to fully predict the gap function of a given instance, based on the field dependence of the DQP, see Fig.~\ref{prediction}.

These findings enable the optimized design of quantum adiabatic protocols with significantly enhanced performance. In the present article, we have presented a practical advantage by feeding  annealing protocols with the information of the DQP, see Figs.~\ref{fidelity_N10} and \ref{fidVsAlpha}.  At least if one disregards the time needed to perform the quench experiments, our procedure leads to a time-to-solution benefit in annealing algorithms which has not been achieved yet, e.g., with pausing \cite{Marshall2019,Chen2020}.

Our study suggests that the annealing gap and the DQP are related in spin glass systems. This enables to draw instance-specific information about the problem Hamiltonian and its energy spectrum in a transverse field from simple quench experiments. Specifically, the magnetization after a quench which is readily measured can reveal the energy gap of the system in a transverse field.  Moreover, it would be interesting to find a relation between the dynamical quantum phase transition and the annealing gap in the thermodynamic limit. This, however, requires a proper scaling of the system size which is not trivial to perform for random Hamiltonians.
In this context, we have also given an example how modern machine learning algorithms can be used to support quantum annealing protocols and thus enhance their efficiency.

This new insight into the annealing gap is expected to trigger further research on optimal annealing protocols and the adiabatic theorem. We already have explicitly demonstrated that partial or full knowledge of the annealing gap can be used to improve the protocol. Interestingly, as our simulation of the protocol defined in Eq.~\eqref{fullGap} has shown, simple adiabatic theorems are not sufficient to determine the \textit{best} protocol, that is, the protocol which for a given annealing time would maximize the final fidelity.

\begin{acknowledgments}
The authors are grateful for enlightening discussions with Gabriel Fern\'andez Fern\'andez, Alexandre Dauphin, and Maciej Lewenstein. The authors acknowledge funding from ``la Caixa'' Foundation (ID 100010434, fellowship code LCF/BQ/PI19/11690013); ERC AdG NOQIA; Agencia Estatal de Investigación (R$\&$D project CEX2019-000910-S, funded by MCIN/ AEI/10.13039/501100011033, Plan National FIDEUA PID2019-106901GB-I00, FPI, QUANTERA MAQS PCI2019-111828-2, Proyectos de I+D+I “Retos Colaboración” RTC2019-007196-7);  Fundació Cellex; Fundació Mir-Puig; Generalitat de Catalunya through the CERCA program, AGAUR Grant No. 2017 SGR 134, QuantumCAT \ U16-011424, co-funded by ERDF Operational Program of Catalonia 2014-2020; EU Horizon 2020 FET-OPEN OPTOLogic (Grant No 899794); National Science Centre, Poland (Symfonia Grant No. 2016/20/W/ST4/00314).
\end{acknowledgments}

\newpage
\clearpage
\ \\
\newpage

\end{document}